\documentclass[reqno,a4paper,10pt,final]{amsart}
\usepackage[T1]{fontenc}
\usepackage[utf8x]{inputenc}
\usepackage{amsmath,amssymb,amstext,mathrsfs,eucal}
\usepackage{array}
\usepackage[vcentermath]{youngtab}
\Yboxdim{4pt}
\usepackage{hyperref}
\hypersetup{%
  pdftitle   = {M2-branes, Einstein manifolds and triple systems},
  pdfkeywords = {M2, M Theory, Bagger-Lambert, Lie algebra, Filippov, 3-algebra, Lie triple system, Chern-Simons, Faulkner, superpotential},
  pdfauthor  = {José Figueroa-O'Farrill},
  pdfcreator = {\LaTeX\ with package \flqq hyperref\frqq}
}
\PrerenderUnicode{éÉśę}
\let\into\hookrightarrow
\newcommand{\half}{\tfrac12}
\newcommand{\fg}{\mathfrak{g}}
\newcommand{\fB}{\mathfrak{B}}
\newcommand{\fF}{\mathfrak{F}}
\newcommand{\fM}{\mathfrak{M}}
\newcommand{\fV}{\mathfrak{V}}
\newcommand{\fk}{\mathfrak{k}}
\newcommand{\fso}{\mathfrak{so}}
\newcommand{\fosp}{\mathfrak{osp}}
\newcommand{\fusp}{\mathfrak{usp}}
\newcommand{\fsp}{\mathfrak{sp}}
\newcommand{\fsu}{\mathfrak{su}}
\newcommand{\fu}{\mathfrak{u}}
\newcommand{\Cl}{\mathrm{C}\ell}
\newcommand{\SO}{\mathrm{SO}}
\newcommand{\SU}{\mathrm{SU}}
\newcommand{\Sp}{\mathrm{Sp}}
\newcommand{\Spin}{\mathrm{Spin}}
\newcommand{\U}{\mathrm{U}}
\newcommand{\RP}{\mathbb{RP}}
\newcommand{\RR}{\mathbb{R}}
\newcommand{\CC}{\mathbb{C}}
\newcommand{\HH}{\mathbb{H}}
\newcommand{\KK}{\mathbb{K}}
\newcommand{\ZZ}{\mathbb{Z}}
\newcommand{\eC}{\mathscr{C}}
\newcommand{\eN}{\mathscr{N}}
\newcommand{\sW}{\mathsf{W}}
\newcommand{\Vbar}{\overline{V}}

\newcommand{\AdS}{\mathrm{AdS}}
\DeclareMathOperator{\Dar}{Rep}
\DeclareMathOperator{\Irr}{Irr}
\DeclareMathOperator{\dvol}{dvol}
\newcommand{\rf}[1]{[\![#1]\!]}
\newcommand{\rh}[1]{(\!(#1)\!)}
\allowdisplaybreaks

\begin{document}
\dedicatory{Poświęcone pamięci Mojego Przyjaciela Krzysztofa Galickiego}
\title{M2-branes, Einstein manifolds and triple systems}
\author{José Miguel Figueroa-O'Farrill}
\address{Institute for the Physics and Mathematics of the Universe, University of Tokyo, Kashiwa, Chiba 277-8586, Japan}
\address{School of Mathematics and Maxwell Institute for Mathematical Sciences, University of Edinburgh, Edinburgh EH9 3JZ, UK}
\begin{abstract}
  This is the written version of a talk given on 1 July 2009 at the XXV Max Born Symposium: the Planck Scale, held in Wrocław, Poland.  I review the possible transverse geometries to supersymmetric M2-brane configurations and discuss the representation-theoretic description of their conjectured dual superconformal Chern--Simons theories.
\end{abstract}
\maketitle

\section{Introduction}
\label{sec:introduction-1}

It is a pleasure to speak at this Max Born Symposium in Wrocław, not just for the obvious reason, but at least on two other accounts: firstly, because I share with Max Born the odd fate of having ended up at the University of Edinburgh (albeit in different departments); and secondly, because my friend Krzysztof Galicki came from Wrocław and studied here before we met as graduate students doing our PhD with Martin Roček in Stony Brook a quarter of a century ago.  It was from him that I learnt, among other things, the proper Polish pronunciation of the city I used to refer to as \emph{Breslavia} and it is to his memory that I dedicate the written version of this talk.

It is not clear whether the subject of my talk is appropriate for a conference on the Planck scale.  It is not a talk about quantum gravity, even though it derives its motivation from an attempt to understand M-theory \cite{Witten:1995ex}, the strong coupling limit of a candidate theory of quantum gravity: namely,  type IIA superstring theory.

We know relatively little about M-theory away from its low-energy limit: eleven-dimensional supergravity \cite{Nahm,CJS}.  The absence of string solutions suggests that this is not a theory of strings, whereas the existence of membrane solutions (so-called M2-branes) suggests that it might be a theory of membranes.  However it is difficult to make this more precise because membranes have resisted every attempt at quantisation.

One way to try to understand at least some aspects of M-theory is via the AdS/CFT correspondence (see, e.g., \cite{AdSCFTReview}).  In that context, the M2-branes play a similar role to the D3-branes of type IIB string theory: namely they give rise to a conjectural correspondence between a gravitational theory and a conformal gauge field theory.  In the case of the D3-brane, this is the duality between type IIB string theory on $\AdS_5 \times S^5$, with equal radii of curvature proportional to $N^{\frac14}$, where $N$ is the number of coincident D3-branes, and the maximally supersymmetric four-dimensional $\SU(N)$ Yang--Mills theory.  In the case of the M2-brane, this duality should relate M-theory on $\AdS_4 \times S^7$, with commensurate radii of curvature now proportional to $N^{\frac16}$, and a maximally supersymmetric three-dimensional conformal field theory, which at the time of the original conjecture \cite{Malda} and for almost a decade thence proved elusive.  In fact, even as recently as 2004, doubts were cast as to the existence of a lagrangian description of such a theory \cite{SchwarzCS}.

This all changed with the pioneering work of Bagger and Lambert \cite{BL1,BL2} and Gustavsson \cite{GustavssonAlgM2} who constructed a maximally supersymmetric three-dimensional conformal field theory.  As shown in \cite{VanRaamsdonkBL} (see also \cite{BermanAMM}) the BLG model can be recast as a superconformal Chern--Simons theory with group $\SU(2) \times \SU(2)$ with opposite levels coupled to matter in the bifundamental representation of the group.  The BLG model is the only known (indecomposable) unitary maximally superconformal three-dimensional field theory involving a finite number of fields and it has been argued \cite{LambertTong} to describe two M2-branes at an orbifold $\RR^8/\ZZ_2$.  The theory dual to any number of M2-branes in $\RR^8/\ZZ_k$ was constructed by Aharony, Bergman, Jafferis and Maldacena \cite{MaldacenaBL} as an $\eN{=}6$ superconformal Chern--Simons theory with gauge group $\U(N)_k \times \U(N)_{-k}$.  One expects that for $k=1,2$ that there ought to be supersymmetry enhancement to $\eN{=}8$ and this has recently been demonstrated \cite{Gustavsson:2009pm,Kwon:2009ar} by considering monopole operators.  Similarly the dual theories to M2-branes on some other geometries have been constructed as superconformal Chern--Simons theories with the appropriate amount of supersymmetry the geometry dictates; although it is fair to say that the dual theory to by far most of the possible transverse geometries (especially those with little supersymmetry) has not been identified; although see, for instance, \cite{OoguriPark,JT} for results in that direction.

This state of affairs motivates the desire to establish a more precise dictionary between the possible transverse geometries to supersymmetric M2-brane configurations and superconformal Chern--Simons theories.  To this end one needs to first determine the possible geometries and also the possible theories and this is what I will discuss in this talk.  The task of determining the possible superconformal Chern--Simons theories is best accomplished using the language and tools of representation theory, particularly the theory of unitary representations of a Lie algebra admitting an ad-invariant inner product.  This ``metric'' property of the Lie algebra allows to relate its unitary representations to certain triple systems, which explains \emph{a posteriori} the important rôle played by ternary algebras in the original BLG model.

The talk will thus consist of two parts.  In the first part, departing from the well-known elementary M2-brane solution of eleven-dimensional supergravity, I will review other supersymmetric solutions which are obtained by replacing the euclidean space transverse to the brane by a Ricci-flat cone admitting parallel spinors.  In dimension 8 there is a wealth of such geometries which were reviewed in \cite{AFHS,MorrisonPlesser} and will be recalled here briefly.  In the second part of the talk I will describe the representation-theoretic underpinnings of superconformal Chern--Simons theories along the lines of \cite{SCCS3Algs}, which is the companion paper to \cite{Lie3Algs}, and discuss their classification.  This sets the stage for a more thorough investigation of the dictionary, which is the subject of an ongoing investigation to be reported on elsewhere.  In particular it suggests the tantalising prospect of associating triple systems to a certain Einstein manifolds.  In the meantime we can view the results described here as one more example of how supersymmetry shines its light into mathematical objects such as triple systems and gives us a fresh reason to investigate them further.

\section{Supersymmetric M2-brane geometries}
\label{sec:supersymm-m2-brane}

For the purposes of this talk, eleven-dimensional supergravity is a system of geometric partial differential equations for an eleven-dimensional lorentzian metric $g$ and a closed 4-form $F$.  We will not write these equations down, but simply mention that they admit a two-parameter family of half-supersymmetric solutions describing a stack of $N$ coincident M2-branes \cite{DS2brane}.  Explicitly, we have the following expressions for $g,F$:
\begin{equation}
  \label{eq:M2}
  \begin{aligned}[m]
    g &= H^{-\frac23}\, ds^2(\RR^{2,1}) + H^{\frac13} ds^2(\RR^8)\\
    F &= \dvol(\RR^{2,1}) \wedge dH^{-1},
  \end{aligned}
\end{equation}
where $H$ is a harmonic function on $\RR^8$ which we will take to be maximally symmetric:
\begin{equation}
  \label{eq:H}
  H = \alpha + \frac{\beta}{r^6}~,
\end{equation}
for $\alpha,\beta\in \RR$ not both equal to zero.  The parameter $\beta$ depends linearly on the number $N$ of M2-branes.  For generic values of $\alpha,\beta$, which means $\alpha \beta \neq 0$, this solution preserves one-half of the supersymmetry, but if either $\alpha$ or $\beta$ vanish, supersymmetry is enhanced to maximal.  If $\beta =0$ there are no M2-branes and the solution is isometric to the Minkowski vacuum $\RR^{10,1}$ with zero $F$, whereas if we take $\alpha = 0$, the solution is isometric to $\AdS_4 \times S^7$ with the radii of curvature in a ratio of $1:2$ and with $F$ proportional to the volume form on the $\AdS_4$ \cite{GibbonsTownsend,DuffGibbonsTownsend}.  This latter solution is known as the near-horizon geometry of the M2-branes, since taking $\alpha$ to zero is formally the same as taking $r$ to zero.

Writing the metric on the euclidean transverse space in spherical polar coordinates, suggests a way to generalise this solution.  We replace
\begin{equation}
  ds^2(\RR^8) = dr^2 + r^2 ds^2(S^7)  \qquad\text{by}\qquad dr^2 + r^2 ds^2(M^7),
\end{equation}
where $M$ is a 7-dimensional riemannian manifold.  For the new $g$ (and the old $F$) to be a solution of eleven-dimensional supergravity it is necessary and sufficient for $M$ to be Einstein with unit radius of curvature.  However, if we want the solution to be supersymmetric, then $M$ should be a spin manifold admitting real Killing spinors; namely, there should exist nonzero spinor fields $\psi$ obeying the equation
\begin{equation}
  \nabla_X \psi = \half X \cdot \psi \qquad\text{for all vector fields $X$},
\end{equation}
where $X \cdot \psi$ is the action of the Clifford bundle $\Cl(TM)$ on the spinor bundle.  As observed by Bär \cite{Baer}, this is equivalent to the cone $C(M) = \RR^{+} \times M$, with metric $dr^2 + r^2 ds^2(M)$, admitting parallel spinors.  If $M$ is assumed to be complete, then a theorem of Gallot \cite{Gallot} says that the cone is either flat, in which case $M$ is locally isometric to $S^7$, or else the holonomy of the cone is irreducible.  Wang \cite{Wang} determined the holonomy representations of irreducible riemannian manifolds admitting parallel spinors and in dimension $8$, which is the dimension of the cone over $M$, these are $\Spin(7)$ acting on the spinor representation, $\SU(4) \subset \SO(8)$ and $\Sp(2) \subset SO(8)$.  The last two correspond to Calabi--Yau 4-folds and hyperkähler 8-manifolds, respectively.  Those 7-manifolds whose cones have $\Spin(7)$, $\SU(4)$ and $\Sp(2)$ holonomy belong to well-known classes: weak $G_2$ holonomy, Sasaki-Einstein and 3-Sasakian, respectively.  The corresponding solutions have the interpretation as the near-horizon geometry of M2-branes at a conical singularity in a special holonomy 8-manifold, which near the singularity is described as the cone over one of the 7-manifolds just mentioned.

Let $\eN$ denote the dimension of the space of Killing spinors on $M$, so that the corresponding M2-brane solution will preserve $2\eN$ supercharges.  The positive values that $\eN$ may take are $1,2,3,4,5,6,8$.  For $\eN>3$ the manifolds $M$ are all smooth quotients of $S^7$ by a finite subgroup $\Gamma$ of $\SO(8)$ which leaves invariant an $\eN$-dimensional subspace of chiral spinors.  There are two smooth manifolds with $\eN{=}8$: $S^7$ itself and $\RP^7$, which is the quotient of $S^7$ by the order two subgroup generated by the antipodal map.  For every finite subgroup of $\SU(2)$, there is an embedding in $\SO(8)$ in such a way that the resulting quotient is smooth and has $\eN{=}5$, unless the subgroup is cyclic in which case $\eN{=}6$, if the order is $>2$, and $\eN{=}8$ if the order is 2.  There are cyclic and binary dihedral, octahedraal and icosahedral subgroups of $\SO(8)$ for which the corresponding quotient has $\eN{=}4$.  The classification has recently been finished in \cite{deMedeiros:2009pp}.  The values $\eN{=}1,2,3$ correspond to weak $G_2$ holonomy, Sasaki-Einstein and 3-Sasakian manifolds, respectively, some of which can of course be sphere quotients.  Many classes of such manifolds are discussed in \cite{AFHS,MorrisonPlesser} and references therein.  The book \cite{MR2382957} by Boyer and Galicki contains a wealth of information on Sasakian and 3-Sasakian geometry.  Table~\ref{tab:geometries} summarises the results reviewed so far.

\begin{table}[h!]
  \centering
  \caption{Supersymmetric M2-brane geometries}
  \begin{tabular}{|>{$}c<{$}|c|c|}\hline
    \eN & Cone & 7-dimensional geometry \\\hline
    8 & $\RR^8$, $\RR^8/\ZZ_2$ & $S^7$, $\RP^7$\\
    4,5,6 & $\RR^8/\Gamma$ & $S^7/\Gamma$\\
    3 & hyperkähler & 3-Sasaki\\
    2 & Calabi--Yau & Sasaki-Einstein\\
    1 & $\Spin(7)$ holonomy & weak $G_2$ holonomy\\
    \hline
  \end{tabular}
  \label{tab:geometries}
\end{table}

To every supersymmetric solution of eleven-dimensional supergravity, one can attach a Lie superalgebra which is generated by the Killing spinors.  This is described in complete generality in \cite{FMPHom}; although for the near-horizon geometries of the supersymmetric M2-brane configurations which are the focus of this talk, they were calculated in \cite{JMFKilling} and shown to be isomorphic to the orthosymplectic Lie superalgebra $\fosp(\eN|4)$, as expected from the AdS/CFT correspondence.  Indeed, this correspondence posits that to every M2-brane configuration with a near-horizon geometry of the form $\AdS_4 \times M$ with $M$ admitting an $\eN$-dimensional space of Killing spinors, there corresponds a three-dimensional $\eN$-extended superconformal field theory.  The three-dimensional conformal superalgebras were classified by Nahm \cite{Nahm} and shown to be isomorphic to $\fosp(\eN|4)$, although it is realised differently.  Focusing on the even Lie algebra $\fso(\eN) \oplus \fsp(4,\RR)$, we recognise $\fso(\eN)$ as the generic isometry algebra of the 7-manifold $M$ and also the R-symmetry of the superconformal field theory, whereas $\fsp(4,\RR) \cong \fso(3,2)$ is the isometry algebra of $\AdS_4$ and also the conformal algebra of $\RR^{2,1}$.  Examples of field theories exhibiting this symmetry are the superconformal Chern--Simons theories with matter, which we now begin to describe.

\section{Superconformal Chern--Simons theories with matter}
\label{sec:superc-chern-simons}

The degrees of freedom of a theory dual to a supersymmetric brane configuration usually include some scalars which parametrise the normal bundle to the brane in the spacetime and its fermionic partners which supersymmetry demands.  Unlike in the case of a D3-brane, for an M2-brane the degrees of freedom of the scalars and the fermions already match, whence any gauge fields present in the theory should contribute no new dynamical degrees of freedom.  This forbids a Yang--Mills-like action, since in three dimensions this has propagating degrees of freedom, but it does not forbid a Chern--Simons term whose gauge fields are non-propagating.  The sort of theories we will be considering thus contain the supersymmetric completion of a Chern-Simons term
\begin{equation}
  \label{eq:Chern-Simons}
  \left(A, dA\right) + \tfrac13 \left([A,A],A\right)~,
\end{equation}
where $A$ is a one-form on $\RR^{2,1}$ with values in a Lie algebra $\fg$ and $\left(-,-\right)$ is an ad-invariant inner product on $\fg$.  This turns $\fg$ into a metric Lie algebra.  If $\fg$ is simple, then $\left(-,-\right)$ is a multiple of the Killing form.  Quantum consistency of the Chern--Simons term quantises this multiple, which is then called the \emph{level} of the Chern--Simons theory.  If $\fg$ is not simple, then there will be a larger space of ad-invariant inner products and quantum consistency now selects a lattice in it.

Matter fields live in supermultiplets which contain a scalar, a Majorana fermion and an auxiliary field we shall ignore in this talk.  The action consists of the standard gauge-covariant kinetic terms, Yukawa couplings and a sextic scalar potential, consistent with conformal invariance.  In a manifestly $\eN{=}1$ superspace formulation, there are two kinds of terms in the matter lagrangian: the kinetic terms and a quartic superpotential which, upon integrating out the auxiliary fields, gives rise to a sextic scalar potential and the Yukawa couplings.  The matter fields belong to a unitary representation of $\fg$.  Unitarity of the theory requires the inner product of the matter fields to be positive-definite, but since the Chern--Simons gauge fields do not propagate their inner product $\left(-,-\right)$ need not be positive-definite and in many cases it will be forced to be indefinite.

Matter fields will also transform in a unitary representation of the $\fso(\eN)$ R-symmetry.  It follows from \cite{Nahm} that the supercharges in the conformal superalgebra transform in the fundamental representation of $\fso(\eN)$.  Since the supercharges relate bosons to fermions, it follows that the bosonic R-symmetry representation $\fB$ must appear in the tensor product decomposition $\fV \otimes \fF$ of the vector representation $\fV$ and the fermionic R-symmetry representation $\fF$, and similarly for $\fB$ and $\fF$ interchanged.  The simplest way to achieve this is for $\fB$ and $\fF$ to be spinorial representations of $\fso(\eN$), with the intertwiners $\fV \otimes \fB \to \fF$ and $\fV \otimes \fF \to \fB$ given by Clifford action.  This means that when $\eN$ is odd, bosons and fermions will be in the same representation whereas if $\eN$ is even the fermionic representation will be obtained from the bosonic one by changing the chirality of the spinor representations.  Table~\ref{tab:spinors} summarises the spinor representations for $\eN\leq 8$.  The notation is such that only the types (real, complex or quaternionic) and their dimension are explicitly written down.  I will use the notation $\Delta^{(\eN)}$ or $\Delta^{(\eN)}_\pm$ for the irreducible spinor representations of $\fso(\eN)$ for $\eN$ odd and even, respectively.  The subscripts refer to the chirality, of course.

\begin{table}[h!]
  \centering
  \caption{Spinor representations of $\fso(\eN)$ for $\eN\leq 8$}
  \begin{tabular}{|*{9}{>{$}c<{$}|}}
    \hline
    \eN & 1 & 2 & 3 & 4 & 5 & 6 & 7 & 8\\\hline
    \fso(\eN) &  & \fu(1) & \fsp(1) & \fsp(1) \oplus \fsp(1) & \fsp(2) & \fsu(4) & \fso(7) & \fso(8)\\
    \text{Spinors} & \RR  & \CC & \HH & \HH \oplus \HH & \HH^2 & \CC^4 &  \RR^8 &  \RR^8 \oplus \RR^8\\\hline
  \end{tabular}
  \label{tab:spinors}
\end{table}

Since gauge transformations and supersymmetry commute (we are not talking about a supergravity theory), both bosons and fermions transform under the same representation of the gauge Lie algebra $\fg$: let's call it $\fM$ generically.  Since matter degrees of freedom are fundamentally real, the type of the R-symmetry representation determines the type of $\fM$, which can read off from the table: real if $\eN{=}1,7,8$, complex if $\eN = 2,6$ and quaternionic if $\eN = 3,4,5$.

For the $\eN{=}1,2,3$ theories we can take the matter to be in any real, complex or quaternionic unitary representations, respectively, with the usual proviso that for $\eN{=}2$ we must take fields and their complex conjugates, in effect working with the underlying real representation obtained by restricting scalars from $\CC$ to $\RR$, and that for $\eN{=}3$ fields are subject to the usual symplectic reality condition on the tensor product of two quaternionic representations.  For a good review of these theories see \cite{GYCS}.

Things get more interesting for $\eN\geq 4$ in that the type of representation and unexpectedly perhaps also the gauge symmetry $\fg$ are constrained.  In order to write down these constraints we must first discuss a refinement of the usual unitary representation theory of Lie algebras which is possible when the Lie algebra is metric.  As we now discuss, this leads quite naturally to the subject of three-algebras or triple systems and may explain \emph{a posteriori} the rôle played by such algebraic objects in the early literature on this topic.

\section{Lie-embeddable unitary representations and triple systems}
\label{sec:lie-embedd-unit}

We now discuss special kinds of unitary representations of a metric Lie algebra.  This summarises the results in \cite{Lie3Algs}, which derived its inspiration from \cite{FaulknerIdeals}.

\subsection{Real unitary representations}
\label{sec:real-unit-repr}

Let $\fg$ be a metric Lie algebra as above and let $U$ denote a real unitary representation.  The inner products on $\fg$ and $U$ will be denoted $\left(-,-\right)$ and $\left<-,-\right>$, respectively.  Since we have $\fg$-invariant inner products on both $\fg$ and $U$, we can take the transpose of the representation map $\fg \to \fso(U)$ to obtain a $\fg$-equivariant linear map $T:\Lambda^2 U \to \fg$, which is surjective if $U$ is a faithful representation.   Explicitly, given $u,v \in U$, we define $T(u,v) \in \fg$ by
\begin{equation}
  \label{eq:T}
  \left(T(u,v), X\right) = \left<X\cdot u, v\right>\qquad\text{for all $X \in \fg$.}
\end{equation}
(In indices, if $X_a$ is a basis for $\fg$ and $u_i$ is a basis for $U$, then $X_a \cdot u_i = T_{a i}^j u_j$ and the map $T(u_i,u_j) = T_{ij}^a X_a$, where $T_{ij}^a$ is obtained from $T_{a i}^j$ by raising and lowering indices with the relevant inner product.)  Now consider the tensor $\Omega$ defined by
\begin{equation}
  \label{eq:omega}
  \Omega(u,v,x,y) = \left(T(u,v),T(x,y)\right).
\end{equation}
It follows that $\Omega$ is a $\fg$-invariant tensor in the representation
\begin{equation}
  S^2\Lambda^2 U = \Lambda^4 U \oplus U^{\yng(2,2)}.
\end{equation}
For general $U$, $\Omega$ will have components in both representations, but for some special representations one or the other component will vanish.  Such representations can be described in the language of triple systems or 3-algebras, more precisely \textbf{metric 3-Leibniz algebras}, as described, for example in \cite{JMF3Defs}.

Define a trilinear product $U \times U \times U \to U$ by
\begin{equation}
\label{eq:3bracket}
  [u,v,w] := T(u,v) \cdot w,
\end{equation}
in terms of which the tensor $\Omega$ can be rewritten as $\Omega(u,v,x,y) = \left<[u,v,x],y\right>$.  The $\fg$-equivariance of $T$ translates into a \emph{fundamental identity} for the triple product:
\begin{equation}
  [x,y,[v,w,z]] = [[x,y,v],w,z] + [v,[x,y,w],z] + [v,w,[x,y,z]],
\end{equation}
whereas the product obeys symmetry properties which follow from the tensorial decomposition of $\Omega$.  These 3-algebras can trace their origin to \cite{FaulknerIdeals} and to \cite{CherSaem} in the present context.

If $\Omega \in \Lambda^4 U$, then $[u,v,w]$ is totally skew-symmetric and defines on $U$ the structure of a \textbf{3-Lie algebra}, a structure formalised by Filippov \cite{Filippov} but going back to the work of Nambu \cite{Nambu:1973qe}.  As conjectured (in a separate context and using a slightly different language) in \cite{FOPPluecker}, there is a unique positive-definite irreducible representation $U$: it is four-dimensional and $\fg = \fsu(2) \oplus \fsu(2)$ with invariant inner product given by the Killing form on one of the simple ideals and the negative of the Killing form on the other.  There exist at least four independent proofs of this fact: a geometric proof based on prolongations of Lie algebras \cite{NagykLie}, two similar proofs based on a combinatorial analysis of the equation \cite{GP3Lie,GG3Lie} and a structural proof \cite{Lor3Lie} based on the classification theorem for simple 3-Lie algebras \cite{LingSimple}.  This 3-Lie algebra is precisely the one in the original work of Bagger--Lambert and Gustavsson and the corresponding Chern--Simons theory is the unique interacting, manifestly unitary, maximally superconformal such theory.

The other extreme case is where $\Omega \in U^{\yng(2,2)}$ is an algebraic curvature tensor.  The triple product now satisfies
\begin{equation}
  \label{eq:Lie}
  [u,v,w] + [v,w,u] + [w,u,v] = 0,
\end{equation}
making $U$ into a \textbf{Lie triple system}.  Lie triple systems are linear approximations to symmetric spaces, in the same way that Lie algebras are linear approximations to Lie groups, and in fact the tensor $\Omega$ in this case is the curvature tensor of the symmetric space.  The classification of positive-definite Lie triple systems is classical and goes back to Cartan's classification of riemannian symmetric spaces.  Such representations $U$ can be used to construct $\eN{=}1$ superconformal theories, but as far as I know these theories are not any more special than the ones constructed out of a generic real representation $U$.

\subsection{Complex unitary representations}
\label{sec:compl-unit-repr}

Now let $V$ be a complex unitary representation of $\fg$ and let $\left<-,-\right>$ now denote a hermitian inner product on $V$, complex antilinear in the second slot in my conventions.  The transpose of the $\fg$ action on $V$ now defines a bilinear map $T: V \times \Vbar \to \fg_\CC$ to the complexification of $\fg$, where $\Vbar$ is the conjugate representation to $V$.  We make $\fg_\CC$ into a complex metric Lie algebra by extending both the Lie bracket and the inner product \emph{complex-bilinearly}.  If $u,v \in V$ we define $T(u,v) \in \fg_\CC$ by \eqref{eq:T}, but where $X \in \fg_\CC$ now.  The tensor $\Omega$, defined in \eqref{eq:omega}, now belongs to the $\fg$-invariants in
\begin{equation}
  S^2(V \otimes \Vbar) = (S^2V \otimes S^2\Vbar) \oplus (\Lambda^2V \otimes \Lambda^2\Vbar).
\end{equation}

The sesquibilinear triple product $V \times \Vbar \times V \to V$ defined by \eqref{eq:3bracket} obeys the following version of the fundamental identity
\begin{equation}
  [x,y,[v,w,z]] = [[x,y,v],w,z] - [v,[y,x,w],z] + [v,w,[x,y,z]],
\end{equation}
where the change in the middle term in due to $\overline{T(x,y)} = - T(y,x)$.  The two extremes, when one or the other component of $\Omega$ vanishes, correspond to representations $V$ where
\begin{equation}
  \label{eq:Jordan}
  [u,v,w] = \pm [w,v,u].
\end{equation}
The positive sign gives rise to \textbf{Jordan triple systems} and the negative to \textbf{anti Jordan triple systems} \cite{FaulknerFerrarAJP}.  Jordan triple systems are in bijective correspondence with hermitian symmetric spaces, and again in this case $\Omega$ is the curvature tensor of the relevant Kähler metric.  Their classification is therefore again classical.  The classification of positive-definite anti Jordan triple system reduces, as we will see, to the classification of certain complex Lie superalgebras \cite{Palmkvist,JMFSimplicity}.  Anti Jordan triple systems are precisely the 3-algebras put forward in \cite{BL4} to reformulate the $\eN{=}6$ theories of \cite{MaldacenaBL}.  Again one can use Jordan triple systems to construct $\eN{=}2$ theories, but to my knowledge they are not more special than the $\eN{=}2$ theories built out of generic $V$.

\subsection{Quaternionic unitary representations}
\label{sec:quat-unit-repr}

Finally we come to the case of quaternionic unitary representations.  The nonexistence of quaternionic Lie algebras means that it is more convenient to think of these representations as complex unitary representations with an invariant quaternionic structure map, denoted $J$.  Hence let $W$ be a complex unitary representation with hermitian structure $\left<-,-\right>$ and let $J: W \to W$ be a $\fg$-equivariant complex antilinear map, obeying $J^2 = -1$ and compatible with the hermitian structure in that \begin{equation}
  \omega(u,v) = \left<u,Jv\right>
\end{equation}
is a $\fg$-invariant complex symplectic structure.  Defining the transpose of the $\fg$-action but relative to $\omega$ instead, we obtain a map $T: S^2W \to \fg_\CC$ and a tensor $\Omega$ defined as in \eqref{eq:omega} which is $\fg$-invariant in the representation
\begin{equation}
  S^2(S^2 W) = S^4 W \oplus W^{\yng(2,2)}.
\end{equation}
$\Omega$ belongs to $W^{\yng(2,2)}$, the triple product defined by \eqref{eq:3bracket} satisfies equation \eqref{eq:Lie}, except that now $[u,v,w] = [v,u,w]$ so we have an \textbf{anti Lie triple system}.  These are the representations for the $\eN{=}5$ theories of \cite{3Lee, BHRSS} and the $\eN{=}4$ theories of \cite{GaiottoWitten} and also appear as building blocks for the $\eN{=}4$ theories of \cite{pre3Lee}.  The classification of positive-definite representations again reduces to the classification of certain Lie superalgebras \cite{JMFSimplicity}.  The representations where $\Omega$ is totally symmetric correspond to hyperkähler symmetric spaces, but there are no nontrivial representations in positive-definite signature and hence no unitary theories based on them.

This discussion is summarised in Table~\ref{tab:matter-reps}, which employs the following notation.  For $\eN$ odd the scalars and fermions live in the same representation, whereas for $\eN$ even they live in representations with opposite chirality for the R-symmetry spinors, and both representations are written, with the top sign corresponding to the scalars and the bottom sign to the fermions, in some conventions.  The notation $\Dar(\fg,\KK)$ denotes the category of positive-definite unitary representations of $\fg$ of type $\KK = \RR, \CC, \HH$, and $\Dar(\fg,\KK)_{\eC}$ the set ot those which are of class $\eC$, where $\eC$ can be either aLTS, aJTS or 3LA for anti Lie triple systems, anti Jordan triple systems or 3-Lie algebras, respectively.  The notation $\Irr(\fg,\KK)_{\eC}$ denotes the subsets of irreducible objects.

\begin{table}[h!]
  \centering
  \caption{Matter representations for $\eN$-extended supersymmetry}
  \begin{tabular}{|>{$}c<{$}|>{$}c<{$}|>{$}c<{$}|}
    \hline
    \eN & \text{Matter representation} & \text{Remarks}\\\hline
    1 & U & U \in \Dar(\fg,\RR)\\
    2 & \Delta^{(2)}_\pm \otimes V \oplus  \Delta^{(2)}_\mp \otimes \Vbar & V \in \Dar(\fg,\CC)\\
    3 & \Delta^{(3)} \otimes W & W \in \Dar(\fg,\HH)\\
    4 & \Delta^{(4)}_\pm \otimes W_1 \oplus  \Delta^{(4)}_\mp \otimes W_2 & W_{1,2} \in \Dar(\fg,\HH)_{\text{aLTS}}\\
    5 & \Delta^{(5)} \otimes W & W \in \Irr(\fg,\HH)_{\text{aLTS}}\\
    6 & \Delta^{(6)}_\pm \otimes V \oplus  \Delta^{(6)}_\mp \otimes \Vbar & V \in \Irr(\fg,\CC)_{\text{aJTS}}\\
    7 & \Delta^{(7)} \otimes U & U \in \Irr(\fg,\RR)_{\text{3LA}}\\
    8 & \Delta^{(8)}_\pm \otimes U & U \in \Irr(\fg,\RR)_{\text{3LA}}\\
    \hline
  \end{tabular}
  \label{tab:matter-reps}
\end{table}

The irreducibility conditions correspond to the notion of an indecomposable theory; namely, one which does not decouple into two or more nontrivial non-interacting theories.  For $\eN<4$ indecomposability does not imply irreducibility (e.g., take $\fg$ simple), whereas for $\eN>4$  indecomposability does imply irreducibility.  For $\eN{=}4$ if $W_1=0$ then $W_2$ has to be irreducible and viceversa; otherwise, indecomposability imposes connectedness of  the corresponding quiver.

\subsection{Embedding Lie (super)algebras}
\label{sec:embedd-lie-super}

One peculiar property of the special representations described above is that they embed in Lie (super)algebras, which means that the triple product in the corresponding triple system is given by nesting two Lie brackets.  For the Lie and Jordan triple systems this is of course a classical result.

Indeed, if $U$ is a Lie triple system, then on the $\ZZ_2$-graded vector space $\fk = \fg \oplus U$, with $\fg$ in degree 0 and $U$ in degree 1, we may define a Lie algebra structure extending that of $\fg$ and the action of $\fg$ on $U$ by declaring $[u,v] = T(u,v)$, for $u,v\in U$.  (Recall that $T$ is skewsymmetric, so this bracket is trying to define a Lie algebra.)  Most of the components of the Jacobi identity are immediate except for one, which is equivalent to \eqref{eq:Lie}.  The resulting $\ZZ_2$-graded Lie algebra $\fk$ is metric by using the inner products on $\fg$ and $U$.  The symmetric space associated to this representation is $K/G$, where $K$ is a Lie group with Lie algebra $\fk$ and $G$ is the closed subgroup with Lie algebra $\fg$.
Similarly, if $V$ is a Jordan triple system, then we consider the complex graded vector space $\fk = V \oplus \fg_\CC \oplus \Vbar$, in degrees $-1,0,1$.  Then we define a Lie algebra structure on $\fk$ in such a way that $\fg_\CC$ is a Lie subalgebra by extending the actions of $\fg_\CC$ on $V$ and $\Vbar$ by $[u,v] = T(u,v)$ for $u\in V$ and $v \in \Vbar$.  Again the only nontrivial component of the Jacobi identity is equation \eqref{eq:Jordan} with the plus sign.

There are similar results for the three classes of representations demanded by $\eN\geq 4$ supersymmetry.  An anti Lie triple systems embeds in a complex Lie  superalgebra $\fg_\CC \oplus W$, whereas an anti Jordan triple system embeds in a complex 3-graded Lie superalgebra $V \oplus \fg_\CC \oplus \Vbar$.  The situation mimics the case of Lie and Jordan triple systems, except that the symmetry of the anti Lie triple system says that $\fg_\CC \oplus W$ is a Lie \emph{super}algebra, while the negative sign in \eqref{eq:Jordan} is (one component of) the Jacobi identity only for a Lie \emph{super}algebra.  Finally, metric 3-Lie algebras embed in a real 3-graded Lie superalgebra $U \oplus \fg \oplus U$.

For positive-definite unitary representations, irreducibility implies (with a minor exception in the case of Lie triple systems) the simplicity of the embedding Lie superalgebra.  Hence this allows one to reduce the classification of positive-definite irreducible representations to extant classifications of simple Lie (super)algebras.  In this way one recovers the classifications of $\eN\geq 5$ superconformal Chern--Simons theories in \cite{3Lee,BHRSS,SchnablTachikawa} from conceptually clear representation-theoretic results.

\subsection{Supersymmetry enhancement}
\label{sec:supersymm-enhanc}

The representation theory also explains the conditions for supersymmetry enhancement.  By studying the decomposition of the R-symmetry spinor representations as a result of the embedding of the R-symmetry Lie algebras $\fso(\eN-1) \into \fso(\eN)$, we can read off the conditions which are required for supersymmetry enhancement.  Table~\ref{tab:susy-enhancement} summarises the decomposition of the matter representations from $\eN$- to ($\eN-1$)-extended supersymmetry.
The notation in the table is such that $U_\CC$ is the complexification of a real representation $U$, whereas $V_\HH$ is the quaternionification of a complex representation $V$ and $\rh{W}$ is a complex representation obtained from a quaternionic representation $W$ by forgetting the quaternionic structure.  As usual, square brackets denote the underlying real representation, so that if $V$ is a complex representation with a real structure, then $[V]_\CC \cong V$.

\begin{table}[h!]
  \centering
  \caption{Decomposition of matter representations}
  \begin{tabular}{|>{$}c<{$}|>{$}c<{$}|>{$}c<{$}|}
    \hline
    \eN & \eN-\text{matter representation} & (\eN-1)-\text{matter representation}\\\hline
    8 & \Delta^{(8)}_+ \otimes U & \Delta^{(7)} \otimes U\\
    7 & \Delta^{(7)}\otimes U & [ (\Delta_+^{(6)} \oplus \Delta_-^{(6)}) \otimes U_\CC] \\
    6 & \Delta_+^{(6)} \otimes V \oplus \Delta_-^{(6)} \otimes \Vbar & \Delta^{(5)} \otimes V_\HH\\
    5 & \Delta^{(5)} \otimes W & \Delta_+^{(4)} \otimes W \oplus \Delta_-^{(4)} \otimes W\\
    4 & \Delta_+^{(4)} \otimes W_1 \oplus \Delta_-^{(4)} \otimes W_2 & \Delta^{(3)} \otimes (W_1 \oplus W_2)\\
    3 & \Delta^{(3)} \otimes W & (\Delta_+^{(2)} \oplus \Delta_-^{(2)})\otimes \rh{W}\\\hline
  \end{tabular}
  \label{tab:susy-enhancement}
\end{table}

We may understand the following supersymmetry enhancements, by looking at the $\eN$-extended matter representation in terms of the ($\eN-1$)-extended representation and then comparing with the generic ($\eN-1$)-extended representation:
\begin{itemize}
\item in $\eN{=}4$, $W_1,W_2 \in \Dar(\fg,\HH)_{\text{aLTS}}$ and the enhancement to $\eN{=}5$ occurs precisely when $W_1 = W_2$;
\item in $\eN{=}5$, $W \in \Irr(\fg,\HH)_{\text{aLTS}}$ and the enhancement to $\eN{=}6$ occurs when $W = V_\HH$, for $V \in \Irr(\fg,\CC)_{\text{aJTS}}$; and
\item finally, in $\eN{=}6$, $V \in \Irr(\fg,\CC)_{\text{aJTS}}$ and enhancement to $\eN{=}7$ occurs when $V=U_\CC$ for $U \in \Irr(\fg,\RR)_{\text{3LA}}$.
\end{itemize}

These enhancements are consistent with relations between the different triple systems; namely, a complex representation $V$ is an anti Jordan triple system if and only if its quaternionification is an anti Lie triple system; and a real representation is a 3-Lie algebra if and only if its complexification is an anti Jordan triple system, whereas if the underlying real representation $\rf{V}$ of a complex representation $V$ is a 3-Lie algebra, then $V$ is an anti Jordan triple system.  Some of these results seem to be new and are described in \cite[Appendix~A]{SCCS3Algs}.

Finally we remark that it follows form Table~\ref{tab:susy-enhancement} that enhancement from $\eN{=}7$ to $\eN{=}8$ does not constrain the representation further, suggesting that $\eN{=}7$ implies $\eN{=}8$.  This is indeed the case, as proved in \cite{SCCS3Algs} by a detailed study of the superpotentials.

\subsection{Superpotentials}
\label{sec:superpotentials}

It is often convenient to write the superconformal Chern--Simons theories in an off-shell formalism in which one of the supersymmetries is manifest.  This is done by working in an $\eN{=}1$ superspace where one of the supercharges acts as a supertranslation.  The choice of supercharge breaks the $\fso(\eN$) R-symmetry to the $\fso(\eN-1)$ stabilizer of the supercharge.  In this formalism the theory is determined by a quartic, gauge-invariant superpotential which is inert under the global $\fso(\eN-1)$ symmetry.  The off-shell superfield $\Xi$ that describes the matter content can always be assembled into the representation of $\fso(\eN-1) \oplus \fg$ appearing in the third column of Table~\ref{tab:susy-enhancement}.  The superpotentials can all be expressed as the superspace integral $\tfrac{1}{16} \int d^2 \theta \, \sW ( \Xi )$, where $\sW$ is a real, quartic, ($\fso(\eN-1) \oplus \fg$)-invariant function.  For all $\eN\geq 4$ the expression for this function is given in Table~\ref{tab:superpotentials}.  In \cite{SCCS3Algs} one can find the expression also for $\eN{=}2,3$.
In the table, the tensor $\Theta$ appearing in the $\eN{=}6$ row is the $\fso(5) \cong \fusp(4)$-invariant symplectic form on $\Delta^{(5)}$ while in the $\eN{=}8$ row it denotes the $\fso(7)$-invariant self-dual Cayley 4-form on $\Delta^{(7)}$.  Repeated indices are contracted with respect to the hermitian inner product on $\Delta^{(\eN-1)}$.

\begin{table}[h!]
  \caption{Superpotentials}
  \centering
  \begin{tabular}{|>{$}c<{$}|>{$}c<{$}|}
    \hline
    \eN & \sW ( \Xi ) \\\hline
    8 & \tfrac{1}{3} \, \Theta_{abcd} \, ( T ( \Xi^a , \Xi^b ) , T ( \Xi^c , \Xi^d ) ) \\
    6 & ( T ( \Xi^a , \Xi^b ) , T ( \Xi^b , \Xi^a ) ) + \Theta_{ab} \, \Theta^{cd} \, ( T ( \Xi^a , \Xi^c ) , T ( \Xi^b , \Xi^d ) ) \\
    5 & -\tfrac{1}{6} \, ( T ( \Xi^\alpha , \Xi^\beta ) , T ( \Xi^\beta , \Xi^\alpha ) ) -\tfrac{1}{6} \, ( T ( \Xi^{\dot \alpha} , \Xi^{\dot \beta} ) , T ( \Xi^{\dot \beta} , \Xi^{\dot \alpha} ) ) + ( T ( \Xi^\alpha , \Xi^{\dot \beta} ) , T ( \Xi^{\dot \beta} , \Xi^\alpha ) ) \\
    4 & \tfrac{1}{6} \, ( T_1 ( \Xi^a , \Xi^b ) , T_1 ( \Xi^b , \Xi^a ) ) + \tfrac{1}{6} \, ( T_2 ( \Xi^a , \Xi^b ) , T_2 ( \Xi^b , \Xi^a ) ) - ( T_1 ( \Xi^a , \Xi^b ) , T_2 ( \Xi^b , \Xi^a ) )\\\hline
  \end{tabular}
  \label{tab:superpotentials}
\end{table}

Finally let me remark that the rigidity of the $\eN\geq 3$ theories translates, using the AdS/CFT correspondence, to a rigidity of 3-Sasakian manifolds and it is has indeed been shown by Pedersen and Poon \cite{MR1605929} (see \cite[Theorem~13.3.24]{MR2382957}) that complete 3-Sasakian manifolds are infinitesimally rigid, a result which came \emph{after} the AdS/CFT correspondence.  Had the dual theories to M2-branes been understood earlier, this would have provided a nice mathematical conjecture which I'm sure Krzysztof would have appreciated.

\section*{Acknowledgments}

It is a pleasure to thank Remigiusz Durka and Jerzy Kowalski-Glikman for the invitation to participate in this meeting and the IPMU for support.  This work was supported by a World Premier International Research Center Initiative (WPI Initiative), MEXT, Japan.  It is a pleasure to thank Hitoshi Murayama for the invitation to visit IPMU and the Leverhulme Trust for the award of a Research Fellowship relieving me of my duties at the University of Edinburgh.  Last, but by no means least, I would like to thank Paul de Medeiros, Elena Méndez-Escobar and Patricia Ritter for a most enjoyable collaboration leading up to the results described in this talk, and Roger Bielawski for the Polish lesson.

\bibliographystyle{utphys}
\bibliography{Sugra,Geometry,Algebra}

\providecommand{\href}[2]{#2}\begingroup\raggedright\begin{thebibliography}{10}

\bibitem{Witten:1995ex}
E.~Witten, ``{String theory dynamics in various dimensions},'' {\em Nucl.
  Phys.} {\bf B443} (1995) 85--126,
\href{http://arxiv.org/abs/hep-th/9503124}{{\tt arXiv:hep-th/9503124}}.

\bibitem{Nahm}
W.~Nahm, ``Supersymmetries and their representations,'' {\em Nucl. Phys.} {\bf
  B135} (1978) 149--166.

\bibitem{CJS}
E.~Cremmer, B.~Julia, and J.~Scherk, ``Supergravity in eleven dimensions,''
  {\em Phys. Lett.} {\bf 76B} (1978) 409--412.

\bibitem{AdSCFTReview}
O.~Aharony, S.~S. Gubser, J.~M. Maldacena, H.~Ooguri, and Y.~Oz, ``Large {N}
  field theories, string theory and gravity,'' {\em Phys. Rep.} {\bf 323}
  (2000) 183--386, \href{http://arxiv.org/abs/hep-th/9905111}{{\tt
  arXiv:hep-th/9905111}}.

\bibitem{Malda}
J.~M. Maldacena, ``The large {$N$} limit of superconformal field theories and
  supergravity,'' {\em Adv. Theor. Math. Phys.} {\bf 2} (1998) 231--252,
  \href{http://arxiv.org/abs/hep-th/9711200}{{\tt arXiv:hep-th/9711200}}.

\bibitem{SchwarzCS}
J.~H. Schwarz, ``{Superconformal Chern-Simons theories},'' {\em JHEP} {\bf 11}
  (2004) 078,
\href{http://arxiv.org/abs/hep-th/0411077}{{\tt arXiv:hep-th/0411077}}.

\bibitem{BL1}
J.~Bagger and N.~Lambert, ``{Modeling multiple M2's},'' {\em Phys. Rev.} {\bf
  D75} (2007) 045020,
\href{http://arxiv.org/abs/hep-th/0611108}{{\tt arXiv:hep-th/0611108}}.

\bibitem{BL2}
J.~Bagger and N.~Lambert, ``{Gauge symmetry and supersymmetry of multiple
  M2-branes},'' {\em Phys. Rev.} {\bf D77} (2008) 065008,
\href{http://arxiv.org/abs/0711.0955}{{\tt arXiv:0711.0955 [hep-th]}}.

\bibitem{GustavssonAlgM2}
A.~Gustavsson, ``{Algebraic structures on parallel M2-branes},'' {\em Nucl.
  Phys.} {\bf B811} (2009) 66--76,
\href{http://arxiv.org/abs/0709.1260}{{\tt arXiv:0709.1260 [hep-th]}}.

\bibitem{VanRaamsdonkBL}
M.~Van~Raamsdonk, ``{Comments on the Bagger-Lambert theory and multiple M2-
  branes},''
\href{http://arxiv.org/abs/0803.3803}{{\tt arXiv:0803.3803 [hep-th]}}.

\bibitem{BermanAMM}
D.~S. Berman, L.~C. Tadrowski, and D.~C. Thompson, ``{Aspects of Multiple
  Membranes},''
\href{http://arxiv.org/abs/0803.3611}{{\tt arXiv:0803.3611 [hep-th]}}.

\bibitem{LambertTong}
N.~Lambert and D.~Tong, ``{Membranes on an orbifold},''
\href{http://arxiv.org/abs/0804.1114}{{\tt arXiv:0804.1114 [hep-th]}}.

\bibitem{MaldacenaBL}
O.~Aharony, O.~Bergman, D.~L. Jafferis, and J.~Maldacena, ``{N=6 superconformal
  Chern-Simons-matter theories, M2-branes and their gravity duals},'' {\em
  JHEP} {\bf 10} (2008) 091,
\href{http://arxiv.org/abs/0806.1218}{{\tt arXiv:0806.1218 [hep-th]}}.

\bibitem{Gustavsson:2009pm}
A.~Gustavsson and S.-J. Rey, ``{Enhanced $N=8$ supersymmetry of ABJM theory on
  $\mathbb{R}^8$ and $\mathbb{R}^8/\mathbb{Z}_2$},''
\href{http://arxiv.org/abs/0906.3568}{{\tt arXiv:0906.3568 [hep-th]}}.

\bibitem{Kwon:2009ar}
O.-K. Kwon, P.~Oh, and J.~Sohn, ``{Notes on Supersymmetry Enhancement of ABJM
  Theory},''
\href{http://arxiv.org/abs/0906.4333}{{\tt arXiv:0906.4333 [hep-th]}}.

\bibitem{OoguriPark}
H.~Ooguri and C.-S. Park, ``{Superconformal Chern-Simons Theories and the
  Squashed Seven Sphere},''
\href{http://arxiv.org/abs/0808.0500}{{\tt arXiv:0808.0500 [hep-th]}}.

\bibitem{JT}
D.~L. Jafferis and A.~Tomasiello, ``{A simple class of N=3 gauge/gravity
  duals},''
\href{http://arxiv.org/abs/0808.0864}{{\tt arXiv:0808.0864 [hep-th]}}.

\bibitem{AFHS}
B.~S. Acharya, J.~M. Figueroa-O'Farrill, C.~M. Hull, and B.~Spence, ``Branes at
  conical singularities and holography,'' {\em Adv. Theor. Math. Phys.} {\bf 2}
  (1998) 1249--1286, \href{http://arxiv.org/abs/hep-th/9808014}{{\tt
  arXiv:hep-th/9808014}}.

\bibitem{MorrisonPlesser}
D.~R. Morrison and M.~R. Plesser, ``Non-spherical horizons, {I},'' {\em Adv.
  Theor. Math. Phys.} {\bf 3} (1999) 1--81,
  \href{http://arxiv.org/abs/hep-th/9810201}{{\tt arXiv:hep-th/9810201}}.

\bibitem{SCCS3Algs}
P.~de~Medeiros, J.~Figueroa-O'Farrill, and E.~Méndez-Escobar,
  ``{Superpotentials for superconformal Chern--Simons theories from
  representation theory},'' \href{http://arxiv.org/abs/0908.2125}{{\tt
  arXiv:0908.2125 [hep-th]}}.

\bibitem{Lie3Algs}
P.~de~Medeiros, J.~Figueroa-O'Farrill, E.~Méndez-Escobar, and P.~Ritter, ``{On
  the Lie-algebraic origin of metric 3-algebras},'' {\em Comm. Math. Phys.}
  {\bf 290} (2009) 871--902, \href{http://arxiv.org/abs/0809.1086}{{\tt
  arXiv:0809.1086 [hep-th]}}.

\bibitem{DS2brane}
M.~Duff and K.~Stelle, ``Multi-membrane solutions of {$D{=}11$} supergravity,''
  {\em Phys. Lett.} {\bf 253B} (1991) 113--118.

\bibitem{GibbonsTownsend}
G.~Gibbons and P.~K. Townsend, ``Vacuum interpolation in supergravity via
  $p$-branes,'' {\em Phys. Rev. Lett.} {\bf 71} (1993) 3754--3757,
  \href{http://arxiv.org/abs/hep-th/9302049}{{\tt arXiv:hep-th/9302049}}.

\bibitem{DuffGibbonsTownsend}
M.~Duff, G.~Gibbons, and P.~K. Townsend, ``Macroscopic superstrings as
  interpolating solitons,'' {\em Phys. Lett.} {\bf 332B} (1994) 321--328,
  \href{http://arxiv.org/abs/hep-th/9405124}{{\tt arXiv:hep-th/9405124}}.

\bibitem{Baer}
C.~B{\"a}r, ``Real {K}illing spinors and holonomy,'' {\em Comm. Math. Phys.}
  {\bf 154} (1993) 509--521.

\bibitem{Gallot}
S.~Gallot, ``Equations différentielles caractéristiques de la sphère,'' {\em
  Ann. Sci. École Norm. Sup.} {\bf 12} (1979) 235--267.

\bibitem{Wang}
M.~Wang, ``Parallel spinors and parallel forms,'' {\em Ann. Global Anal. Geom.}
  {\bf 7} (1989), no.~1, 59--68.

\bibitem{deMedeiros:2009pp}
P.~de~Medeiros, J.~Figueroa-O'Farrill, S.~Gadhia, and E.~Méndez-Escobar,
  ``{Half-BPS quotients in M-theory: ADE with a twist},''
\href{http://arxiv.org/abs/0909.0163}{{\tt arXiv:0909.0163 [hep-th]}}.

\bibitem{MR2382957}
C.~P. Boyer and K.~Galicki, {\em Sasakian geometry}.
\newblock Oxford Mathematical Monographs. Oxford University Press, Oxford,
  2008.

\bibitem{FMPHom}
J.~M. Figueroa-O'Farrill, P.~Meessen, and S.~Philip, ``Supersymmetry and
  homogeneity of {M}-theory backgrounds,'' {\em Class. Quant. Grav.} {\bf 22}
  (2005) 207--226, \href{http://arxiv.org/abs/hep-th/0409170}{{\tt
  arXiv:hep-th/0409170}}.

\bibitem{JMFKilling}
J.~M. Figueroa-O'Farrill, ``On the supersymmetries of {A}nti-de~{S}itter
  vacua,'' {\em Class. Quant. Grav.} {\bf 16} (1999) 2043--2055,
  \href{http://arxiv.org/abs/hep-th/9902066}{{\tt arXiv:hep-th/9902066}}.

\bibitem{GYCS}
D.~Gaiotto and X.~Yin, ``{Notes on superconformal Chern-Simons-matter
  theories},'' {\em JHEP} {\bf 08} (2007) 056,
\href{http://arxiv.org/abs/0704.3740}{{\tt arXiv:0704.3740 [hep-th]}}.

\bibitem{FaulknerIdeals}
J.~R. Faulkner, ``On the geometry of inner ideals,'' {\em J. Algebra} {\bf 26}
  (1973) 1--9.

\bibitem{JMF3Defs}
J.~Figueroa-O'Farrill, ``{Deformations of 3-algebras},''
  \href{http://arxiv.org/abs/0903.4871}{{\tt arXiv:0903.4871 [hep-th]}}.

\bibitem{CherSaem}
S.~Cherkis and C.~Sämann, ``{Multiple M2-branes and generalized 3-Lie
  algebras},'' {\em Phys. Rev.} {\bf D78} (2008) 066019,
\href{http://arxiv.org/abs/0807.0808}{{\tt arXiv:0807.0808 [hep-th]}}.

\bibitem{Filippov}
V.~Filippov, ``{$n$}-{L}ie algebras,'' {\em Sibirsk. Mat. Zh.} {\bf 26} (1985),
  no.~6, 126--140, 191.

\bibitem{Nambu:1973qe}
Y.~Nambu, ``{Generalized Hamiltonian dynamics},'' {\em Phys. Rev.} {\bf D7}
  (1973)
2405--2414.

\bibitem{FOPPluecker}
J.~M. Figueroa-O'Farrill and G.~Papadopoulos, ``{P}lücker-type relations for
  orthogonal planes,'' {\em J. Geom. Phys.} {\bf 49} (2004) 294--331,
  \href{http://arxiv.org/abs/0211170}{{\tt arXiv:0211170 [math.AG]}}.

\bibitem{NagykLie}
P.-A. Nagy, ``Prolongations of {L}ie algebras and applications,''
  \href{http://arxiv.org/abs/0712.1398}{{\tt arXiv:0712.1398 [math.DG]}}.

\bibitem{GP3Lie}
G.~Papadopoulos, ``{M2-branes, 3-Lie Algebras and Plucker relations},'' {\em
  JHEP} {\bf 05} (2008) 054,
\href{http://arxiv.org/abs/0804.2662}{{\tt arXiv:0804.2662 [hep-th]}}.

\bibitem{GG3Lie}
J.~P. Gauntlett and J.~B. Gutowski, ``Constraining maximally supersymmetric
  membrane actions,'' \href{http://arxiv.org/abs/0804.3078}{{\tt
  arXiv:0804.3078 [hep-th]}}.

\bibitem{Lor3Lie}
P.~de~Medeiros, J.~Figueroa-O'Farrill, and E.~Méndez-Escobar, ``{Lorentzian
  Lie 3-algebras and their Bagger--Lambert moduli space},'' {\em JHEP} {\bf 07}
  (2008) 111, \href{http://arxiv.org/abs/0805.4363}{{\tt arXiv:0805.4363
  [hep-th]}}.

\bibitem{LingSimple}
W.~X. Ling, {\em On the structure of $n$-{Lie} algebras}.
\newblock PhD thesis, Siegen, 1993.

\bibitem{FaulknerFerrarAJP}
J.~R. Faulkner and J.~C. Ferrar, ``Simple anti-{J}ordan pairs,'' {\em Comm.
  Algebra} {\bf 8} (1980), no.~11, 993--1013.

\bibitem{Palmkvist}
J.~Palmkvist, ``{Three-algebras, triple systems and 3-graded Lie
  superalgebras},'' \href{http://arxiv.org/abs/0905.2468}{{\tt arXiv:0905.2468
  [hep-th]}}.

\bibitem{JMFSimplicity}
J.~Figueroa-O'Farrill, ``{Simplicity in the Faulkner construction},''
  \href{http://arxiv.org/abs/0905.4900}{{\tt arXiv:0905.4900 [hep-th]}}.

\bibitem{BL4}
J.~Bagger and N.~Lambert, ``{Three-Algebras and N=6 Chern-Simons Gauge
  Theories},'' {\em Phys. Rev.} {\bf D79} (2009) 025002,
\href{http://arxiv.org/abs/0807.0163}{{\tt arXiv:0807.0163 [hep-th]}}.

\bibitem{3Lee}
K.~Hosomichi, K.-M. Lee, S.~Lee, S.~Lee, and J.~Park, ``{N=5,6 Superconformal
  Chern-Simons Theories and M2-branes on Orbifolds},'' {\em JHEP} {\bf 09}
  (2008) 002,
\href{http://arxiv.org/abs/0806.4977}{{\tt arXiv:0806.4977 [hep-th]}}.

\bibitem{BHRSS}
E.~A. Bergshoeff, O.~Hohm, D.~Roest, H.~Samtleben, and E.~Sezgin, ``{The
  Superconformal Gaugings in Three Dimensions},'' {\em JHEP} {\bf 09} (2008)
  101,
\href{http://arxiv.org/abs/0807.2841}{{\tt arXiv:0807.2841 [hep-th]}}.

\bibitem{GaiottoWitten}
D.~Gaiotto and E.~Witten, ``{Janus Configurations, Chern-Simons Couplings, And
  The Theta-Angle in N=4 Super Yang-Mills Theory},''
\href{http://arxiv.org/abs/0804.2907}{{\tt arXiv:0804.2907 [hep-th]}}.

\bibitem{pre3Lee}
K.~Hosomichi, K.-M. Lee, S.~Lee, S.~Lee, and J.~Park, ``{N=4 Superconformal
  Chern-Simons Theories with Hyper and Twisted Hyper Multiplets},'' {\em JHEP}
  {\bf 07} (2008) 091,
\href{http://arxiv.org/abs/0805.3662}{{\tt arXiv:0805.3662 [hep-th]}}.

\bibitem{SchnablTachikawa}
M.~Schnabl and Y.~Tachikawa, ``{Classification of N=6 superconformal theories
  of ABJM type},''
\href{http://arxiv.org/abs/0807.1102}{{\tt arXiv:0807.1102 [hep-th]}}.

\bibitem{MR1605929}
H.~Pedersen and Y.~S. Poon, ``A note on rigidity of {$3$}-{S}asakian
  manifolds,'' {\em Proc. Amer. Math. Soc.} {\bf 127} (1999), no.~10,
  3027--3034.

\end{thebibliography}\endgroup

\end{document}